\documentclass[11pt]{article}
\usepackage{comment}
\usepackage[margin=2.5cm]{geometry}
\usepackage{amsmath,amssymb,extarrows,mathtools,graphicx,subfigure,setspace,bbold,physics}
\usepackage{cite}
\usepackage{slashed}
\usepackage[dvipsnames]{xcolor}
\usepackage{hyperref,caption,diagbox}
\hypersetup{colorlinks=true, linkcolor=blue, citecolor=magenta, linktoc=page}
\usepackage{tikz}
\usepackage{color}
\usepackage{arydshln,leftidx,mathtools}
\usepackage{float} 
\usepackage{placeins} 
\usetikzlibrary{decorations.pathreplacing}
\numberwithin{equation}{section}
\newcommand{\be}{\begin{equation}}
	\newcommand{\bea}{\begin{eqnarray}}
		\newcommand{\eea}{\end{eqnarray}}
	\newcommand{\ba}{\begin{align}}
		\newcommand{\ea}{\end{align}}
	\newcommand{\ee}{\end{equation}}

\usepackage{chngcntr}
\begin{document}
\begin{titlepage}
		\thispagestyle{empty}

	\vspace{.4cm}
	\begin{center}
		\noindent{\Large \textbf{Krylov Complexity in Lifshitz-type Dirac Field Theories}}\\
		
		\vspace*{15mm}
		\vspace*{1mm}
		{Hamid R. Imani,  Komeil Babaei Velni, 
  M. Reza Mohammadi Mozaffar
}
		
		\vspace*{1cm}
		
		{\it Department of Physics, University of Guilan, P.O. Box 41335-1914, Rasht, Iran\\}
		
		\vspace*{0.5cm}
    	{E-mails: {\tt \{hamidrezaimani@webmail.,babaeivelni@, mmohammadi@\}guilan.ac.ir}}
		
		\vspace*{1cm}
	\end{center}

\begin{abstract}
We study Krylov complexity in Lifshitz-type Dirac field theories with a generic dynamical critical exponent $z$. By computing the Lanczos coefficients for massless and massive cases, we analyze the growth and saturation behavior of Krylov complexity in different regimes. We incorporate a hard UV cutoff and investigate the effects of lattice discretization, revealing fundamental differences between continuum and lattice models. In the presence of a UV cutoff, Krylov complexity exhibits an initial exponential growth followed by a linear regime, with saturation values of the Lanczos coefficients dictated by the cutoff scale. For the lattice model, we find a fundamental departure from the continuum case: due to the finite Krylov basis, Krylov complexity saturates rather than growing indefinitely. Our findings suggest that Lifshitz scaling influences operator growth and information spreading in quantum systems. We further find that increasing the Lifshitz exponent $z$ suppresses Krylov complexity, entropy, and Lanczos growth in both massless and massive cases, while enhancing K-variance. This trend reverses under a hard UV cutoff, where complexity and entropy increase with $z$. In lattice models, early-time complexity and $b_n$ decay shift with $z$, echoing the continuum behavior of massive and massless regimes.

\end{abstract}

\end{titlepage}

	\newpage
	
	\tableofcontents
	\noindent
	\hrulefill
	
	\onehalfspacing

\section{Introduction}

The study of operator growth in quantum systems has attracted significant attention in recent years due to its relevance to thermalization, quantum chaos and information spreading in quantum field theory (QFT) and many-body physics. Among the most powerful tools to quantify operator growth is the \emph{Krylov complexity} (or \emph{K-complexity})~\cite{Parker:2018yvk}, which measures the spread of an operator in \emph{Krylov space}—an orthonormal basis constructed through repeated commutators of the operator with the Hamiltonian. The dynamics in Krylov space are governed by the Lanczos algorithm, yielding a sequence of \emph{Lanczos coefficients} $\{b_n\}$ that encode the structure of operator evolution and provide insight into growth rates and saturation behavior.

Recent developments have substantially deepened our understanding of K-complexity as a diagnostic of integrability, chaos, and dynamical universality. In particular, K-complexity has been employed to distinguish between integrable and chaotic systems based on the saturation behavior and fluctuations of the Lanczos coefficients~\cite{Rabinovici2022,Ali2019,Alishahiha:2022anw,Bhattacharyya:2023grv,Adhikari:2023evu,Huh:2023jxt,Caputa:2024vrn,Alishahiha:2024rwm,Nandy:2024htc,Camargo:2024deu,Balasubramanian:2024ghv,Das:2024tnw}. However, as emphasized e.g.~in~\cite{Erdmenger2023,Vardian2024}, neither linear growth of Lanczos coefficients nor saturation in K-complexity alone serves as a definitive hallmark of chaos, since these features can also emerge in integrable systems without exhibiting the full signatures of quantum chaos such as random matrix level statistics or eigenstate thermalization. More refined signatures have also been identified in chaotic systems, including rise–ramp–plateau structures in transition amplitudes, exponential early-time growth, and correlations with classical Lyapunov exponents~\cite{Anegawa2024,Avdoshkin2022,Hashimoto2023}. These signatures, which complement earlier diagnostics such as the slope–dip–ramp–plateau structure in spectral form factors and exponential growth of out-of-time-order correlators (OTOCs), have been extended to quantum field theoretic settings where deviations from holographic expectations have been observed. More recently, non-relativistic and thermal field theories have been analyzed using K-complexity diagnostics~\cite{He2024a,He2024b}, revealing temperature-dependent transitions in complexity growth and suppressed delocalization at large dynamical exponent $z$.

Theoretical frameworks suggest that in chaotic quantum systems, the Lanczos coefficients grow asymptotically linearly as $b_n \sim \alpha n + \gamma$, leading to exponential K-complexity growth, $K_O(t) \sim \exp(2\alpha t)$~\cite{Parker:2018yvk}. This behavior has motivated the result that the growth rate $\lambda_K = 2\alpha$ bounds the Lyapunov exponent $\lambda_L$ at infinite temperature, and the conjecture that it continues to do so at finite temperature~\cite{Parker:2018yvk}. The Lyapunov exponent characterizes chaos via the exponential growth of out-of-time-order correlators (OTOCs)~\cite{Maldacena:2016chaos}.

However, exponential growth of K-complexity is also observed in free and integrable QFTs with unbounded power spectra. To distinguish genuine chaotic behavior, recent studies have proposed introducing UV cutoffs or lattice regularization to modify the spectral density and disrupt the universal exponential scaling~\cite{Avdoshkin2022,Camargo:2022,Vasli:2023}. These methods reveal more intricate operator dynamics and highlight the role of finite Hilbert space dimensionality and coarse-graining in shaping Krylov space.

In parallel, there has been growing interest in quantum field theories with non-relativistic scaling symmetry. Such theories, known as \emph{Lifshitz field theories}, exhibit anisotropic scaling
\begin{equation}
	t \rightarrow \lambda^{z} t, \quad x^i \rightarrow \lambda x^i,
\end{equation}
where $z$ is the dynamical critical exponent dictating the relative scaling of time and space. The case $z=1$ corresponds to Lorentz invariance, while $z\neq1$ describes a wide class of physical systems, particularly near quantum critical points in condensed matter and holography. The hallmark of Lifshitz theories is the inclusion of higher-derivative spatial terms in the action. In free fermionic models, a generalized Lifshitz-Dirac Lagrangian in $d$ dimensions is given by~\cite{Vasli:2024Einlifshtz}
\begin{equation}\label{eq:GenericLifshitzDirac}
	\mathcal{L} = \bar{\Psi}\left( i\gamma^0\partial_0 + i\sum_{j=1}^{d} \gamma^j T_{(j)}^{z-1} \partial_j - m^z \right) \Psi,
\end{equation}
where $T_{(j)} = \sqrt{-\partial_j^2}$. In two dimensions, two prototype Lifshitz-Dirac Lagrangians have been proposed (see~\cite{Alexandre:2021,Vasli:2024Einlifshtz} and references therein)
\begin{align}
	\mathcal{L}_{\mathrm{I}} &= \bar{\Psi}\left(i\gamma^0\partial_0 + \gamma^1(i\partial_1)^z - m^z\right)\Psi, \label{LifactionI} \\
	\mathcal{L}_{\mathrm{II}} &= \bar{\Psi}\left(i\gamma^0\partial_0 + i\gamma^1 T^{z-1} \partial_1 - m^z\right)\Psi, \label{LifactionII}
\end{align}
where $T = \sqrt{-\partial_1 \partial^1}$. The first formulation, Eq.~\eqref{LifactionI}, is well-defined for odd $z$, while the second, Eq.~\eqref{LifactionII}, provides a consistent extension to all integer $z \geq 1$ through the use of the spatial operator $T = \sqrt{-\partial_1^2}$. The corresponding dispersion relation takes the form
\begin{equation}
	\epsilon_k^2 = f(k)^{2z} + m^{2z},
\end{equation}
where the function $f(k)$ correspond to different models
\begin{equation}\label{dispersion_relation_bistar}
	f_{\mathrm{I}}(k) = (-\sin k)^z, \quad
	f_{\mathrm{II,\,odd}}(k) = -(\sin k)^z, \quad
	f_{\mathrm{II,\,even}}(k) = -|\sin k|(-\sin k)^{z-1}.
\end{equation}
All formulations reduce to the relativistic case when $z=1$, and exhibit non-relativistic dispersion for $z\neq1$.
Recently, there have been many attempts to investigate different aspects of quantum chaos, entanglement measures and computational complexity in the context of Lifshitz theories, which have led to a remarkably rich and varied range of new insights, e.g.,~\cite{MohammadiMozaffar:2017nri, He:2017wla, Gentle:2017ywk, MohammadiMozaffar:2017chk, MohammadiMozaffar:2018vmk, MohammadiMozaffar:2019gpn, Hartmann:2021vrt, Mozaffar:2021nex, Mintchev:2022xqh, Mintchev:2022yuo,Khoshdooni:2025ddk}. For example, Ref.~\cite{Mozaffar:2021nex} analyzes entanglement structure in Lifshitz-invariant field theories and offers a scalar-field perspective that motivates the fermionic extension explored in~\cite{Vasli:2023}. In Ref.~\cite{Hartmann:2021vrt}, the authors observe vanishing entanglement entropy in certain Lifshitz backgrounds, raising intriguing questions about quantum correlations. Moreover, Ref.~\cite{MohammadiMozaffar:2018vmk} investigates the influence of the Lifshitz exponent on the time evolution of entanglement entropy, providing insight into dynamical scaling effects relevant to our study. While significant progress has been made in understanding K-complexity in bosonic field theories, the role of fermions—especially in non-relativistic settings—remains less explored. Fermionic Lifshitz field theories introduce fundamentally different structures into Krylov dynamics. For instance, due to the anti-commuting nature of fermionic operators, the autocorrelation function is generally not an even function of time. This leads to non-vanishing $a_n$ coefficients in the Lanczos recursion, in contrast to the bosonic case where $a_n = 0$ identically. Moreover, the spectral function for Dirac fields carries a distinct thermal weighting via $1/\cosh(\beta\omega/2)$, differing from the $1/\sinh(\beta\omega/2)$ behavior found in scalar fields. These differences result in a richer structure for Krylov evolution, including alternating patterns in the recursion coefficients and oscillatory features in the Krylov amplitudes due to particle–antiparticle contributions. Studying K-complexity in fermionic Lifshitz theories is also motivated by the broader question of how quantum statistics and anisotropic scaling interplay in shaping operator growth and complexity. Lifshitz scaling with large $z$ suppresses delocalization in bosonic theories~\cite{Vasli:2023}, but it remains unclear whether similar suppression mechanisms or universal behavior emerge in fermionic counterparts. By systematically analyzing Dirac fields in both continuum and lattice settings, our work provides new insights into these questions and opens avenues for contrasting operator spreading in theories with different spin-statistics and scaling symmetries. An important structural feature of fermionic systems is the appearance of nonzero diagonal Lanczos coefficients $a_n$, which are typically absent in scalar field theories. This alternating structure enriches the dynamics in Krylov space and motivates the study of K-complexity in fermionic Lifshitz theories as a complementary probe to the bosonic case. Moreover, in contrast to bosonic fields where Krylov complexity can diverge at high temperatures due to dense mode accumulation, fermionic systems exhibit finite K-complexity. This statistical distinction highlights the role of Fermi–Dirac statistics in shaping operator growth and further motivates the investigation of fermionic models in the Lifshitz context.

In this work, we investigate K-complexity in Lifshitz-type Dirac field theories, both in the continuum and on discretized lattices. Our aim is to compute the Lanczos coefficients in massless and massive regimes, characterize their scaling with the dynamical exponent $z$, and analyze the effects of discretization and UV cutoff on the asymptotic behavior. We show that the interplay of non-Lorentz invariant scaling and fermionic structure gives rise to novel features in operator growth and saturation, distinguishing these systems from scalar field theories~\cite{Vasli:2023}.

The paper is organized as follows. In section~\ref{sec:preliminaries_k_complexity}, we review the basics of Lifshitz field theories and introduce the tools necessary for analyzing K-complexity. Section~\ref{sec:k_complexity_dirac} analyzes the massless and massive Dirac field cases. In section~\ref{sec:lanczos_uv_cutoff}, we explore the effect of UV cutoffs and discretized lattices. We conclude with a summary and outlook in section~\ref{Conclusions}.

\section{Preliminaries: K-complexity in QFT} \label{sec:preliminaries_k_complexity}

To compute K-complexity, one requires either the spectral function or the thermal autocorrelation function. The spectral function encodes how different frequency modes contribute to the system's excitations. By integrating the spectral function over momentum, one obtains the Wightman power spectrum, describing the energy distribution across frequencies.
 The spectral function for a Lifshitz-type Dirac field is given by~\cite{Laine:2016thermal}
\begin{equation}\label{spectralfunction}
	\rho(\omega, k) = \frac{N}{\epsilon_k} (\omega \pm m^z) \left[\delta(\omega - \epsilon_k) - \delta(\omega + \epsilon_k)\right],
\end{equation}
where $z$ is the critical exponent, and the dispersion relation reads
\begin{equation}\label{dispersionrelation}
	\epsilon_k = \sqrt{k^{2z} + m^{2z}}.
\end{equation}
This dispersion relation highlights the anisotropic scaling of the Lifshitz Dirac field. For $z=1$, it reduces to the standard relativistic form, while for $z>1$, higher spatial derivatives dominate, leading to non-relativistic scaling behavior. The Wightman power spectrum is obtained from
\begin{equation}\label{omega}
	\Pi_W(\omega, k) = \frac{1}{\cosh\left( \frac{\beta \omega}{2} \right)} \rho(\omega, k),
\end{equation}
and is integrated over momentum as
\begin{equation}\label{Wightman}
	f_W(\omega) = \int \frac{d^{d-1} k}{(2\pi)^{d-1}} \Pi_W(\omega, k).
\end{equation}
For numerical purposes, one may approximate $\frac{1}{\cosh(\beta\omega/2)}$ by $e^{-\beta|\omega|/2}$ in the limit $\beta\omega\gg1$, which introduces only minor deviations at low $\beta\omega$, as verified through direct comparison.
 Substituting Eq.~\eqref{omega} into Eq.~\eqref{Wightman} and performing the integration yields
\begin{equation}\label{WPS}
	f_W(\omega) = N(m, \beta, d, z) \frac{( \omega \pm m^z ) (\omega^2 - m^{2z})^{\frac{d-1}{2z} - 1}}{z \cosh(\beta \omega/2)} \left[ \theta(\omega - m^z) - \theta(-\omega - m^z) \right],
\end{equation}
where $N(m, \beta, d, z)$ is a normalization factor determined by
\begin{equation}\label{norm}
	\int \frac{d\omega}{2\pi} f_W(\omega) = 1.
\end{equation}
The Krylov basis is denoted by $\{|\mathcal{O}_n)\}$, and the time-evolved operator is expanded as
\begin{equation}\label{AU}
	|\mathcal{O}(t)) = \sum_{n=0}^{\infty} i^n \phi_n(t) |\mathcal{O}_n),
\end{equation}
where the normalization condition reads
\begin{equation}\label{normalization}
	\sum_{n=0}^{\infty} |\phi_n(t)|^2 = 1.
\end{equation}
The amplitudes $\phi_n(t)$ satisfy the discretized Schrödinger equation
\begin{equation}\label{disSchrodinger}
	i \frac{d\phi_n}{dt} = a_n \phi_n + b_{n+1} \phi_{n+1} + b_n \phi_{n-1}, \quad \text{with} \quad \phi_{-1}(t) \equiv 0.
\end{equation}
The initial condition $\phi_0(t)$ is obtained via the inverse Fourier transform of the Wightman power spectrum Eq.~\eqref{WPS}. K-complexity, $K_{\mathcal{O}}(t)$, quantifies the spread of the operator across Krylov space and is defined by
\begin{equation}\label{KC}
	K_{\mathcal{O}}(t) = \sum_{n=0}^{\infty} n |\phi_n(t)|^2.
\end{equation}
A small $K_{\mathcal{O}}(t)$ indicates operator localization within a few Krylov modes, while a large $K_{\mathcal{O}}(t)$ signals significant operator growth. In chaotic systems, $K_{\mathcal{O}}(t)$ typically exhibits exponential growth at early times before eventual saturation~\cite{Parker:2018yvk}. However, similar behaviors have been observed in certain integrable systems~\cite{Avdoshkin2022,Camargo:2022}. To characterize fluctuations around the average complexity, the $k$-th order K-variance is defined as~\cite{Caputa:2021OperatorGrowth}
\begin{equation}\label{Kvariance}
	\delta_{\mathcal{O}}(t) = \frac{\left(\sum_{n} n^k |\phi_n(t)|^2 - K_{\mathcal{O}}(t)^k\right)^{1/k}}{K_{\mathcal{O}}(t)}
\end{equation}
where for $k=2$, it reduces to the standard deviation of Krylov mode occupation. Moreover, the K-entropy, measuring the delocalization of the operator in Krylov space, is given by~\cite{Barbon:2019BeyondScrambling}
\begin{equation}\label{Kentropy}
	S_{\mathcal{O}}(t) = -\sum_{n} |\phi_n(t)|^2 \log |\phi_n(t)|^2.
\end{equation}
A small $S_{\mathcal{O}}(t)$ indicates a peaked distribution, while a large $S_{\mathcal{O}}(t)$ reflects a more uniform spreading across Krylov modes. Further, to compute the Lanczos coefficients $b_n$, one first extracts the moments $\mu_n$ from the power spectrum $f(\omega)$ as
\begin{equation}\label{mu2n}
	\mu_n = \frac{1}{2\pi} \int_{-\infty}^{\infty} d\omega \, \omega^n f(\omega).
\end{equation}
These moments form the basis for constructing the Krylov chain and determining the Lanczos coefficients, which are computed recursively via~\cite{Avdoshkin:1994Recursion}
\begin{align}\label{recursion}
	b_n &= \sqrt{M_n^{(n)}}, \qquad b_{-1} = b_0 \equiv 1, \qquad M_\ell^{(-1)} = 0, \qquad M_\ell^{(0)} = (-1)^\ell \mu_\ell, \nonumber\\
	a_n &= -\frac{M_{n+1}^{(n)}}{b_n^2} + \frac{M_n^{(n-1)}}{b_{n-1}^2}, \qquad L_\ell^{(0)} = (-1)^{\ell+1} \mu_{\ell+1}, \nonumber\\
	M_\ell^{(j)} &= L_\ell^{(j-1)} + a_{j-1} \frac{M_{\ell-2}^{(j-1)}}{b_{j-1}^2}, \qquad
	L_\ell^{(j)} = \frac{M_{\ell+1}^{(j)}}{b_j^2} - \frac{M_\ell^{(j-1)}}{b_{j-1}^2}, \qquad \ell = j,\ldots,n.
\end{align}

Here, $M_\ell^{(j)}$ and $L_\ell^{(j)}$ represent iterative coefficients built from the moments $\mu_n$. The Lanczos coefficients $\{a_n, b_n\}$ thus fully determine the Krylov basis and the time evolution of complexity, making the recursive Lanczos algorithm a powerful tool for analyzing operator growth.

\section{K-Complexity in the Continuum Limit} \label{sec:k_complexity_dirac}

\subsection{Massless case}
\begin{figure}[H]
	\centering
	\includegraphics[width=0.48\linewidth]{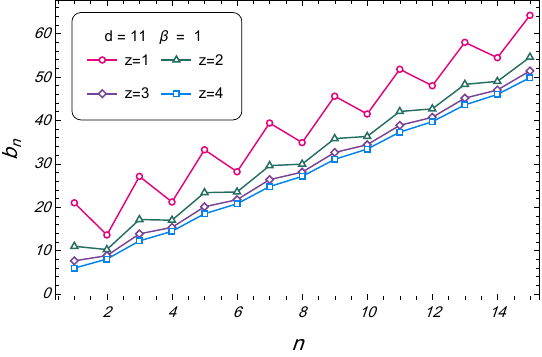}
	\hspace*{0.2cm}
	\includegraphics[width=0.48\linewidth]{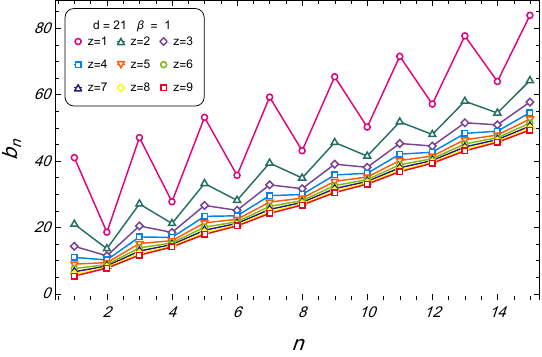}
	\caption{Lanczos coefficients in the massless regime for various dynamical exponents $z$ and spacetime dimensions $d$.}
	\label{Lanczos coefficients in the massles}
\end{figure}

\begin{figure}[H]
	\centering
	\includegraphics[width=0.48\linewidth]{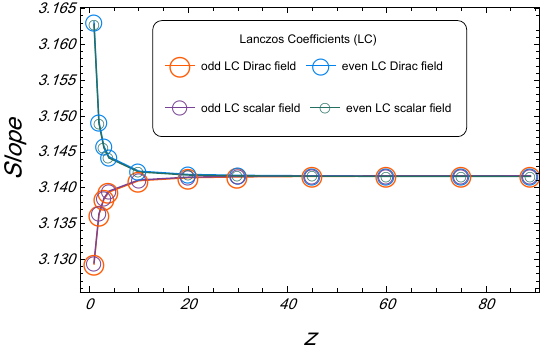}
	\caption{Slope of the Lanczos coefficients in $d=90$ for fermionic and bosonic fields, illustrating dependence on $z$.}
	\label{fig:slopeferbos}
\end{figure}

To derive the massless power spectrum of the Dirac field, we begin with Eq.~\eqref{WPS}. The exponent governing the power spectrum depends on the parameters $d$ and $z$. We assume the minimal dimension $d=2$ and the smallest dynamical exponent $z=1$. Under these conditions, the lowest achievable exponent is $-1/2$, leading to the constraint $d-1\geq z$. The massless Dirac field power spectrum takes the form
\begin{equation}
	f^{W}(\omega) = -\frac{\pi \left(\frac{\beta \omega}{2}\right)^{\xi} \left[\theta(-\omega) - \theta(\omega)\right] e^{-\frac{\beta |\omega|}{2}} }{\omega \Gamma\left(\frac{d-1}{z}\right)},
\end{equation}
where $\xi = \frac{d-1}{z}$. The corresponding moments are given by
\begin{equation}
	\mu_k =
	\begin{cases}
		\displaystyle \frac{2^{2n}\,\beta^{-2n}\,\Gamma(\xi+2n)}{\Gamma(\xi)}, & k=2n, \\[2mm]
		0, & k=2n+1,
	\end{cases}
	\label{miomassless_in_term_of_xi_even_odd}
\end{equation}
showing that only even moments are nonzero. This behavior is analogous to the bosonic Lifshitz field, where moments are even in both the massless and massive cases~\cite{Vasli:2023}. However, as we will show, the massive Dirac field exhibits both even and odd moments. In fermionic cases, where the autocorrelation function is not an even function~\cite{Avdoshkin2022}, the Lanczos coefficients include both $b_n$ and $a_n$. Our numerical results indicate that\footnote{The $\pm$ sign in $a_n$ originates from the $\pm$ sign in the autocorrelation function of the Dirac field, reflecting contributions from particles and antiparticles. However, the final results are independent of particle type.}
\begin{equation}
	a_n = \mp (-1)^n m^z.
\end{equation}
By contrast, in bosonic fields, $a_n=0$ in both massless and massive cases. The alternating structure in $a_n$, proportional to $m^z$, reflects intrinsic features of fermionic fields and introduces additional complexity in Krylov space evolution.
As shown in Fig.~\ref{Lanczos coefficients in the massles}, the $y$-intercepts of the Lanczos coefficients $b_n$ decrease as the dynamical exponent $z$ increases, indicating slower initial growth of K-complexity with stronger anisotropy. Moreover, $b_n$ increases with dimension $d$, suggesting faster operator growth in higher dimensions. According to our numerical calculations, the slope of Lanczos coefficients in both fermionic and bosonic fields depends on $d$ and $z$. As shown in Fig.~\ref{fig:slopeferbos}, when $\xi \equiv \frac{d-1}{z}$ approaches one, the slope converges to $\pi$. This behavior is not tied to a physically special point, but rather reflects specific combinations of spacetime dimension $d$ and dynamical exponent $z$ for which the spectral properties mimic those in~\cite{Camargo:2022}. The resulting slope of $\pi$ is consistent with prior findings in models with continuous spectra and is connected to exponential operator growth and quantum chaos~\cite{Maldacena:2016chaos}.

\begin{figure}[h!]
	\centering
	\includegraphics[width=0.45\textwidth]{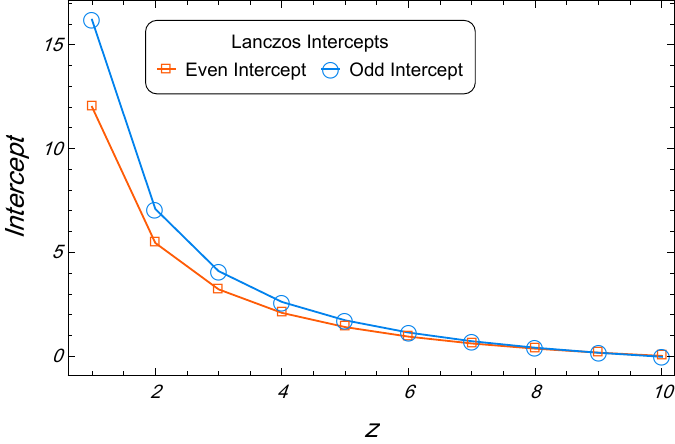}
	\includegraphics[width=0.45\textwidth]{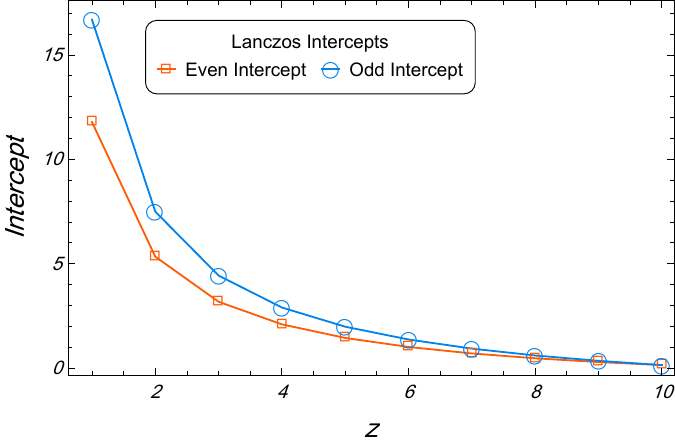}
	\caption{\footnotesize Left: Intercept $\gamma$ of $b_n$ vs.\ $z$ in massless Dirac field. Right: same for massive Dirac field. $\gamma$ decreases rapidly with $z$, suppressing early-time growth.}
	\label{fig:yintercept}
\end{figure}

We further analyze how Lifshitz scaling controls the Krylov exponent by explicitly fitting the Lanczos coefficients to the linear form $b_n \sim \alpha n + \gamma$ in both massless and massive Dirac field theories. Our results show that the slope $\alpha$ systematically converges to $\pi$ as $z$ increases, while the intercept $\gamma$ decreases sharply, as illustrated in Figures~\ref{fig:yintercept}. These trends persist across both even and odd sectors and indicate that increasing the dynamical exponent suppresses early-time growth while fixing the long-time exponential behavior. This provides a quantitative account of how Lifshitz anisotropy modulates the structure of operator spreading in Krylov space.

Next, we investigate the behavior of K-complexity at $d=21$ for various $z$ values in fermionic and bosonic fields and compare their evolution. This comparison highlights distinct features of operator growth in fermionic systems.

\begin{figure}[H]
	\centering
	\includegraphics[width=0.48\linewidth]{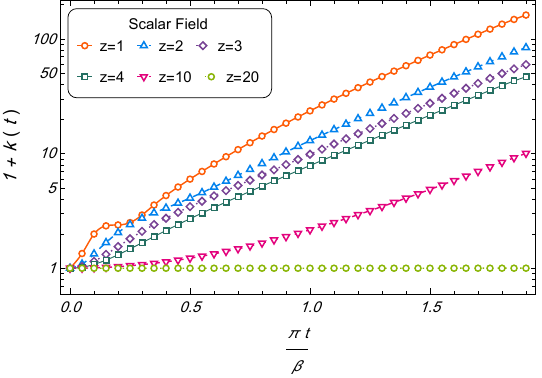}
	\hspace*{0.2cm}
	\includegraphics[width=0.48\linewidth]{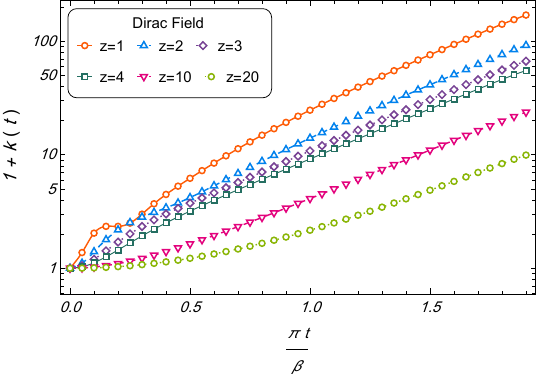}
	\caption{Evolution of K-complexity in massless bosonic and Dirac fields for various $z$ values in $d=21$.}
	\label{fig:fandbcomplexity}
\end{figure}

The two-point functions of free massless Dirac and scalar fields in Lifshitz-type theories, after applying the scaling parameter, are
\begin{equation}\label{Diractf_scaled}
	\phi_0^{\text{Dirac}}(t) = \left(\frac{4t^2}{\beta^2} + 1\right)^{-\xi/2} \cos\left( \xi \tan^{-1}\frac{2t}{\beta} \right),
\end{equation}
\begin{equation}\label{bosontf}
	\phi_0^{\text{Scalar}}(t) = \left(\frac{4t^2}{\beta^2} + 1\right)^{-(\xi-1)/2} \cos\left( (\xi-1) \tan^{-1}\frac{2t}{\beta} \right).
\end{equation}
Our results, shown in Fig.~\ref{fig:fandbcomplexity}, indicate that for $d=21$ and $z=1$, K-complexity is similar between fermionic and bosonic fields. However, as $\xi$ approaches one, differences become more pronounced.

\begin{figure}[h!]
	\centering
	\includegraphics[width=0.5\textwidth]{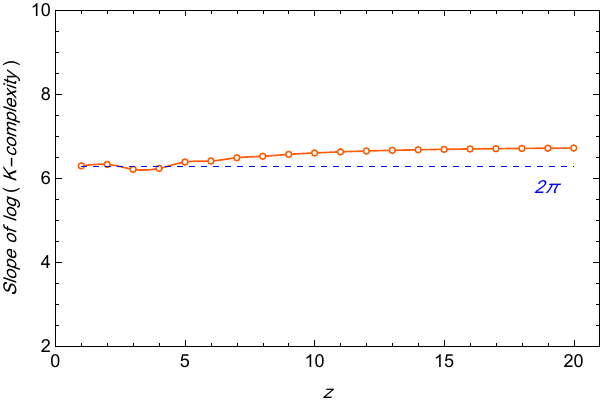}
	\caption{Slope of $\log K(t)$ versus dynamical exponent $z$ in the massless Dirac field at $m = 0$ and $d = 21$, extracted over the time interval $t \in [1.5, 2]$. The blue dashed line indicates the asymptotic limit $\lambda_K = 2\pi$.}
	\label{fig:logslope_vs_z}
\end{figure}
In the massless regime, both Dirac and scalar autocorrelation functions simplify significantly in the zero-temperature limit $\beta \to \infty$. For Dirac fields (Eq.~\ref{Diractf_scaled}), $\phi_0(t) \to 1$, implying $\phi_{n>0}(t) \to 0$ and thus $K(t) \to 0$. A similar behavior is observed for scalars (Eq.~\ref{bosontf}). Interestingly, for $\xi = 1$, the scalar autocorrelation remains exactly constant at all $\beta$, leading to vanishing K-complexity at all temperatures. These results emphasize that thermal excitations are crucial for nontrivial operator growth in both scalar and fermionic Lifshitz-type field theories. To further quantify the effect of the dynamical exponent, we numerically extract the slope of $\log K(t)$ in the interval $t \in [1.5, 2]$ for the massless Dirac field with various values of $z$. The results, shown in Fig.~\ref{fig:logslope_vs_z}, indicate that although the early-time growth of K-complexity is suppressed at larger $z$, the extracted slope asymptotically approaches the universal value $2\pi$, consistent with exponential operator growth. Notably, for large $z$, the fitted slope in the chosen interval slightly exceeds $2\pi$. This deviation is expected, as the exponential growth regime is delayed for higher $z$, and the time interval may still lie within the transient phase. To fully capture the asymptotic rate, one would need to evaluate the slope over a later time interval beyond the initial delay.

\begin{figure}[H]
	\centering
	\includegraphics[width=0.48\linewidth]{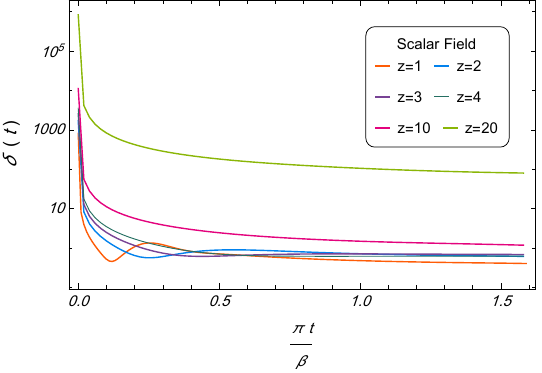}
	\hspace*{0.2cm}
	\includegraphics[width=0.48\linewidth]{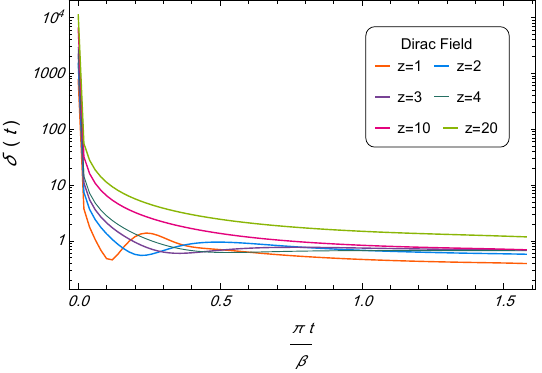}
	\caption{K-variance in massless Dirac and scalar fields for various $z$ values in $d=21$.}
	\label{fig:KVKEm0}
\end{figure}

The K-variance computed via Eq.~\eqref{Kvariance} is shown in Fig.~\ref{fig:KVKEm0} which shows this quantity generally decreases over time, reflecting how operator fluctuations around the average complexity diminish. Larger $z$ values lead to higher variance, indicating slower operator spreading in highly anisotropic systems. No significant difference between fermionic and bosonic fields is observed in this behavior.

\begin{figure}[H]
	\centering
	\includegraphics[width=0.48\linewidth]{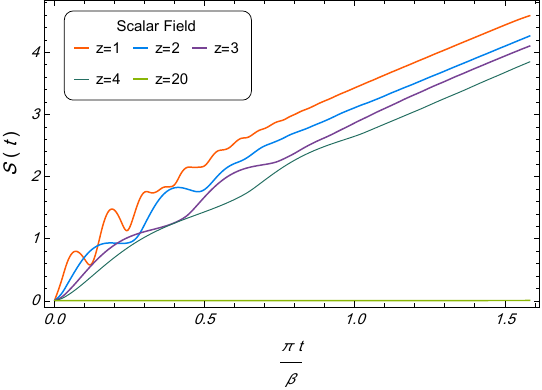}
	\hspace*{0.2cm}
	\includegraphics[width=0.48\linewidth]{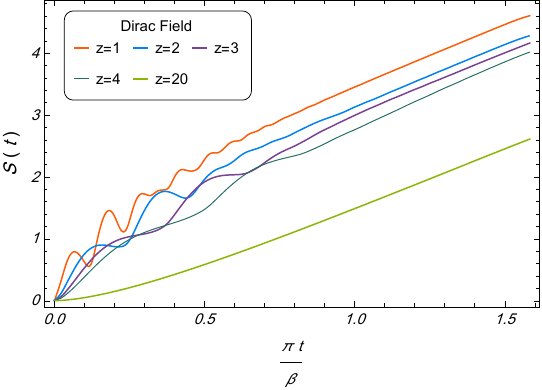}
	\caption{K-entropy in massless Dirac and scalar fields as a function of time for various $z$ values in $d=21$.}
	\label{fig:KEm0}
\end{figure}
Fig~\ref{fig:KEm0} presents the operator delocalization characterized by the K-entropy (see Eq.~\eqref{Kentropy}). Our numerical results indicate that this quantity generally increases with time, marking progressive operator growth. For larger $z$, K-entropy growth slows, consistent with suppressed operator spreading. Remarkably, for bosonic fields at $z=20$, K-entropy remains exactly zero, matching the behavior of K-complexity and reflecting operator freezing in Krylov space at high anisotropy.

\subsection{Massive case}\label{largemass}
In this section, we extend the previous analysis to the massive case. The power spectrum for a massive Dirac fermion in a Lifshitz-type theory is given by
\begin{equation}\label{massivespectralfunction}
	f^W(\omega) = \frac{\pi^{3/2} \cdot 2^{-\xi} \cdot e^{-\frac{\beta |\omega|}{2}} \cdot m^{-\frac{z(\xi + 1)}{2}} \cdot \beta^{\frac{\xi - 1}{2}} \left(\omega^2 - m^{2z} \right)^{\xi/2} \left[ \theta(\omega - m^z) - \theta(-m^z - \omega) \right]}{ \Gamma\left( \frac{\xi}{2} \right) (m^z + \omega) K_{\frac{\xi + 1}{2}}\left( \frac{m^z \beta}{2} \right)}.
\end{equation}
The presence of the mass term $m^z$ introduces a spectral gap, as enforced by the Heaviside functions $\theta(\omega - m^z)$ and $\theta(-m^z - \omega)$, restricting $|\omega|>m^z$. To facilitate the analysis, we derive expressions for both even and odd moments. Compact expressions are obtained using shorthand functions introduced in Appendix~\ref{appendix:calculations}. The even moments are given by
\begin{equation}\label{miok}
	\mu_{\text{even}} = \frac{\pi^2\,2^{-\frac{d+3z+1}{z}}\,m^{\frac{1}{2}(-d-z+1)}\,\beta^{-\frac{d+(4n+1)z-1}{2z}}}
	{K_{\frac{d+z-1}{2z}}\left(\frac{m^z\beta}{2}\right)}
	\sum_{j=1}^4 A_j^{(\text{even})}(n,d,z,m,\beta),
\end{equation}
while the odd moments are
\begin{equation}
	\mu_{\text{odd}} = \frac{\pi^2\,2^{-\frac{d+3z+1}{z}}\,m^{\frac{1}{2}(-d-z+1)}\,\beta^{-\frac{d+(4n+3)z-1}{2z}}}
	{K_{\frac{d+z-1}{2z}}\left(\frac{m^z\beta}{2}\right)}
	\sum_{j=1}^3 A_j^{(\text{odd})}(n,d,z,m,\beta).
	\label{mioodd}
\end{equation}
\footnotetext{We have expressed Eq.~\eqref{mioodd} in terms of $\xi$ in Appendix~\ref{appendix:calculations}.}
These expressions demonstrate that, similar to the massless case, \textit{i.e.}, Eq.~\eqref{miomassless_in_term_of_xi_even_odd}, the moments exhibit universal scaling controlled by $\xi$. Throughout, we substitute $m^z\to\tilde{m}$ without loss of generality, maintaining $m^z$ fixed under changes in $z$.
\begin{figure}[H]
	\centering
	\includegraphics[width=0.45\linewidth]{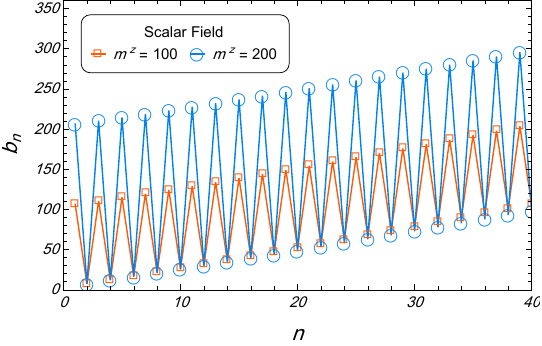}
	\hspace*{.01cm}
	\includegraphics[width=0.45\linewidth]{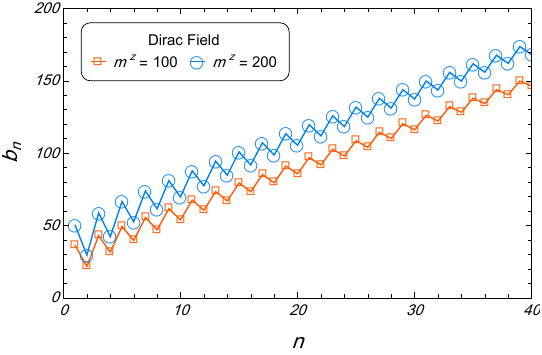}
	\caption{Lanczos coefficients for Dirac and scalar fields at fixed $d=13$, $z=2$, and various masses. The mass increases from left to right, leading to enhanced staggering between even and odd modes.}
	\label{fig:fermionbosonm1002001000040000z2d13}
\end{figure}

Figure~\ref{fig:fermionbosonm1002001000040000z2d13} illustrates the effect of mass on the separation of even and odd Lanczos coefficients in Dirac and scalar fields at fixed $d$ and $z$. The increasing staggering of K-complexity with larger mass reflects enhanced asymmetry between even and odd Krylov amplitudes. We observe that while mass-induced staggering occurs in both fields~\cite{Camargo:2022}, it is substantially more pronounced for bosonic fields, suggesting that mass affects bosonic Krylov dynamics more strongly. We now compute K-complexity for the massive Dirac field. The autocorrelation function $\phi_0(t)$ is obtained from Eq.~\eqref{massivespectralfunction} as
\begin{equation}\label{phi0_equation}
	\begin{split}
		\phi_0(t) &= \frac{1}{2\sqrt{\beta}\,K_{\frac{1+\xi}{2}}\left(\frac{\tilde{m}\beta}{2}\right)}
		\left(\frac{\beta}{4t^2+\beta^2}\right)^{\xi/2} \\[2mm]
		&\quad \times \Biggl\{
		\sqrt{-2it+\beta}\,(2it+\beta)^{\xi/2}
		\Bigl[K_{\frac{-1+\xi}{2}}\left(\frac{\tilde{m}(-2it+\beta)}{2}\right)
		+ K_{\frac{1+\xi}{2}}\left(\frac{\tilde{m}(-2it+\beta)}{2}\right)\Bigr] \\[2mm]
		&\qquad -\, (-2it+\beta)^{\xi/2}\sqrt{2it+\beta}
		\Bigl[K_{\frac{-1+\xi}{2}}\left(\frac{\tilde{m}(2it+\beta)}{2}\right)
		- K_{\frac{1+\xi}{2}}\left(\frac{\tilde{m}(2it+\beta)}{2}\right)\Bigr]
		\Biggr\}.
	\end{split}
\end{equation}

\begin{figure}[H]
	\centering
	\includegraphics[width=0.45\linewidth]{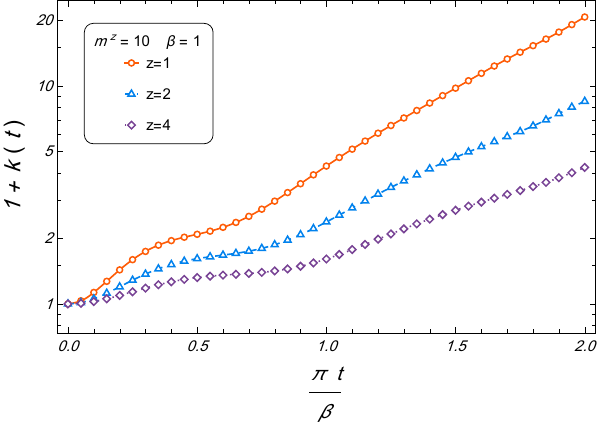}
	\hspace*{.01cm}
	\includegraphics[width=0.45\linewidth]{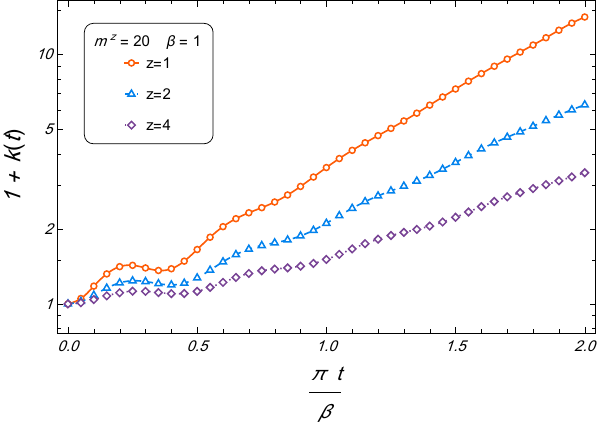}
	\caption{K-complexity in the large mass limit for different values of $m$ and $z$ with $d=5$.}
	\label{fig:km1020z12}
\end{figure}

\begin{figure}[H]
	\centering
	\includegraphics[width=0.45\linewidth]{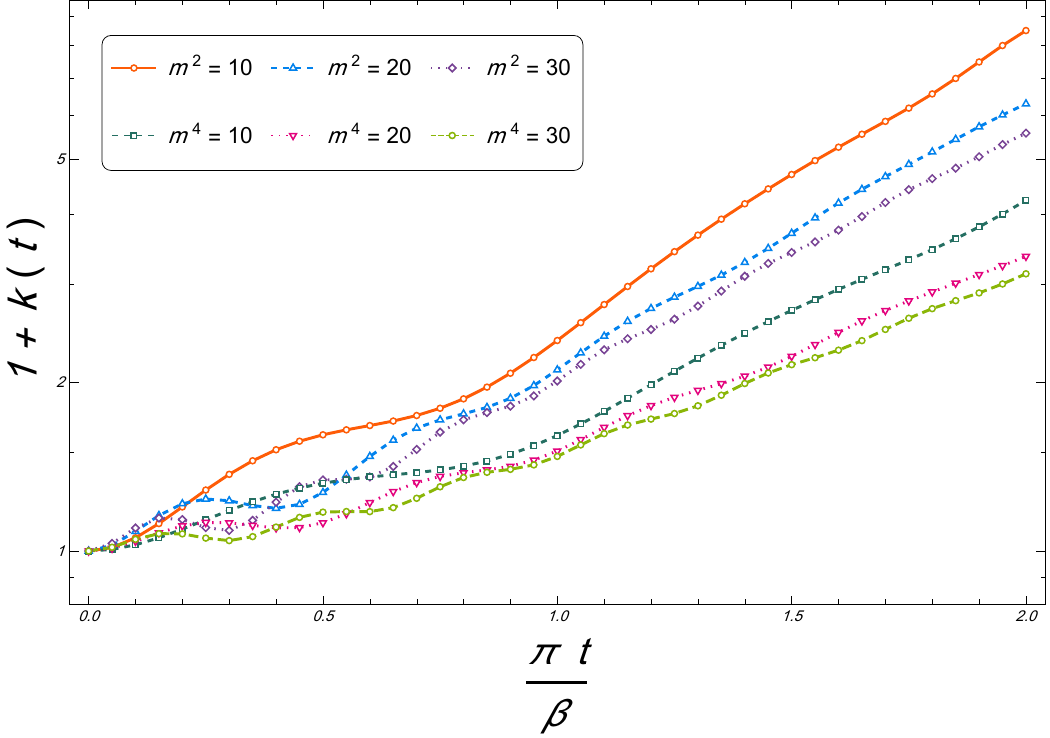}
	\caption{Evolution of K-complexity for several values of mass and dynamical exponent.}
	\label{fig:km102030z2}
\end{figure}

\begin{figure}[H]
	\centering
	\includegraphics[width=0.45\linewidth]{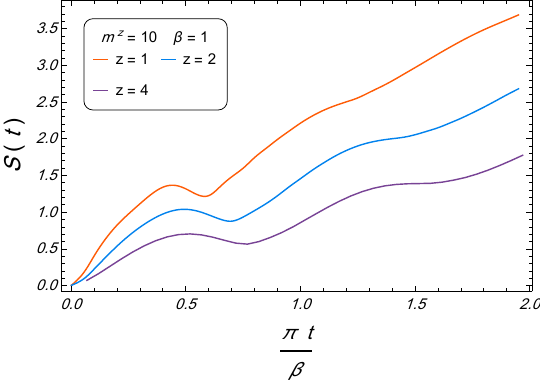}
	\hspace*{.01cm}
	\includegraphics[width=0.45\linewidth]{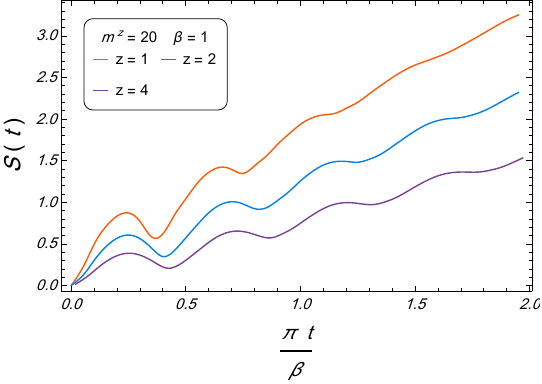}
	\caption{K-entropy for massive Dirac fields for several values of $m$ and $z$ with $d=5$.}
	\label{fig:kem1020z12}
\end{figure}

\begin{figure}[H]
	\centering
	\includegraphics[width=0.45\linewidth]{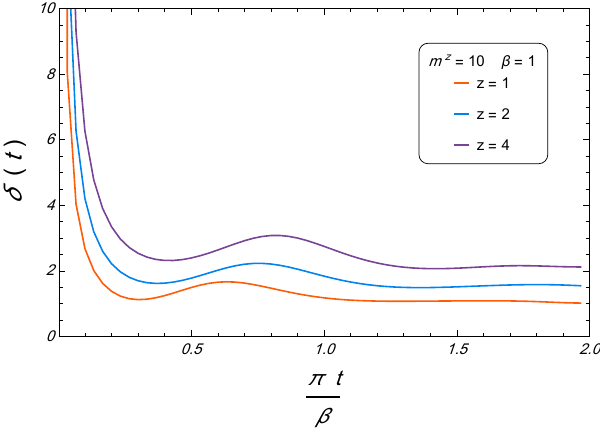}
	\hspace*{.01cm}
	\includegraphics[width=0.45\linewidth]{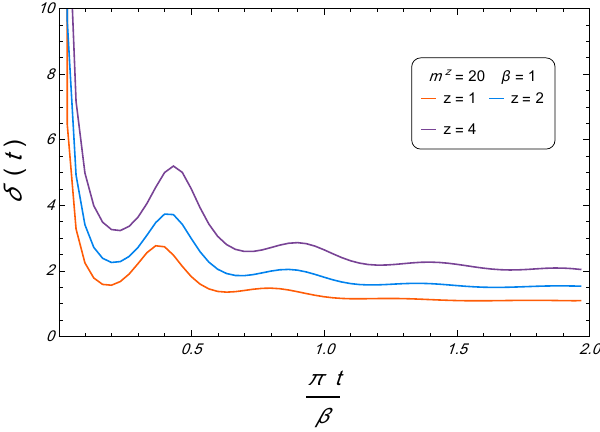}
	\caption{K-variance for different values of $m$ and $z$ with $d=5$.}
	\label{fig:kv1020z12}
\end{figure}

We compute K-complexity, K-entropy, and K-variance numerically for various masses, dimensions, and dynamical exponents. The corresponding results are shown in Figs.~\ref{fig:km1020z12}, \ref{fig:km102030z2}, \ref{fig:kem1020z12} and \ref{fig:kv1020z12}.

Figure \ref{fig:km1020z12} reveals that K-complexity growth slows with increasing mass, indicating that larger masses suppress Krylov space exploration. A higher $z$ further reduces the growth rate. In the zero-temperature limit $\beta\to\infty$, the autocorrelation function simplifies to $\phi_0(t)=e^{i t m^z}$, implying complete localization in the first Krylov basis state and vanishing complexity, $K(t)\to 0$. This matches the physical expectation that without thermal excitations, the system remains in its ground state without nontrivial operator growth.

As illustrated in Fig.~\ref{fig:km102030z2}, smaller masses lead to rapid early-time complexity growth (nearly quadratic), transitioning into a linear regime at later times.\footnote{We use logarithmic plots for K-complexity except in section~\ref{sec:lanczos_uv_cutoff} to better illustrate behavior across time.} Increasing mass suppresses both early growth and the linear slope, consistent with a larger energy gap impeding operator spread. In the massive regime, the behavior of the Krylov growth rate is further altered: the slope of $\log K(t)$ becomes smaller than the universal value $2\pi$, reflecting the suppression of operator spreading due to the spectral gap introduced by the mass term. This trend is consistent with previous observations in the literature (see e.g., \cite{Camargo:2022}), where increasing the mass reduces the Krylov growth rate. Our numerical results for fixed mass $m^z = 10$ further reveal that increasing the dynamical exponent $z$ leads to an additional reduction in the slope. The slopes were extracted by fitting $\log K(t)$ in the interval $t \in [1.5, 2]$, where the exponential regime becomes dominant. As shown in Fig.~\ref{fig:km1020z12}, the extracted values of the slope are $\lambda_K \approx \pi \times \{1.59514, 1.40298, 1.26399\}$ for $z = \{1, 2, 4\}$, respectively. This indicates that both increasing the mass and increasing the Lifshitz exponent act to suppress the early-time exponential growth rate of K-complexity. These findings are in agreement with the behavior observed in Fig.~\ref{fig:km102030z2}, and highlight the combined effect of mass and anisotropy in slowing down operator growth in Lifshitz-type Dirac field theories.

The time evolution of K-entropy is presented in Fig.~\ref{fig:kem1020z12}. As mass increases, the growth of K-entropy becomes slower, reflecting reduced randomization due to the spectral gap. Larger $z$ values similarly lead to lower K-entropy, indicating stronger suppression of operator delocalization under increased anisotropy.

Finally, Fig.~\ref{fig:kv1020z12} shows the behavior of the K-variance. A higher mass extends the early-time domain of fluctuating K-variance, indicating stronger localization, while a larger $z$ increases the overall K-variance, suggesting a delayed uniform exploration across the Krylov space.

\section{Lanczos Coefficients and K-complexity in the Presence of UV Cutoff} \label{sec:lanczos_uv_cutoff}

\subsection{K-complexity with hard UV cutoff}

In this section, we investigate the effects of introducing a hard UV cutoff into the Wightman power spectrum. Specifically, we modify the integration domain of the power spectrum to introduce an upper limit at $\Lambda^z$
\begin{equation}
	f^W(\omega) = \frac{\omega - m^z}{z \cosh \left( \frac{\beta \omega}{2} \right)} \int_{m^z}^{\Lambda^z} \left( \delta(\omega - \epsilon) - \delta(\epsilon + \omega) \right) \left( \epsilon^2 - m^{2z} \right)^{\frac{\xi - 2}{2}} d\epsilon.
\end{equation}
In the regime $1\ll \beta m^z\ll \beta \Lambda^z$, the above expression simplifies to
\begin{equation}
	f^{W}(\omega) = \mathcal{N}(m,\beta,\xi)\,
	\frac{e^{-\frac{\beta|\omega|}{2}} \left(\omega^2 - m^{2z}\right)^{\xi/2}}{z \left(m^z + \omega\right)}
	\left[ \theta(\omega - m^z)\, \theta(\Lambda^z - \omega)
	- \theta(-\omega - m^z)\, \theta(\Lambda^z + \omega) \right],
\end{equation}
which captures the essential behavior of the power spectrum at low temperatures. Here, the exponential factor $e^{-\beta |\omega|/2}$ ensures thermal suppression at high frequencies, while the step functions enforce the mass and cutoff bounds. The moments $\{\mu_{n}\}$ can be readily obtained from this modified spectrum. In particular, for $z=2$ and $d=5$, the moments are
\begin{equation}\label{eq:mu_n}
	\mu_n = \frac{(2i)^k\,\beta^{-k}\,e^{\frac{\beta}{2}(\Lambda^2 + m^2)}}{D}
	\left[ 2\cos\left(\frac{\pi k}{2}\right)\,\Delta\Gamma_{k+2}
	+ i\beta m\, \sin\left(\frac{\pi k}{2}\right)\,\Delta\Gamma_{k+1} \right],
\end{equation}
where
\[
D = e^{\frac{\beta\Lambda^2}{2}}(\beta m^2 + 2) - e^{\frac{\beta m^2}{2}}(\beta\Lambda^2 + 2),
\]
and
\[
\Delta\Gamma_{k+\alpha} = \Gamma\left(k+\alpha,\,\frac{m^2\beta}{2}\right) - \Gamma\left(k+\alpha,\,\frac{\beta\Lambda^2}{2}\right).
\]
The introduction of the UV cutoff modifies both the power spectrum and its moments, constraining the operator's Krylov space dynamics. The thermal suppression, mass corrections, and step function restrictions collectively shape the behavior of operator growth under finite energy limits.

\begin{figure}[H]
	\centering
	\includegraphics[width=0.49\linewidth]{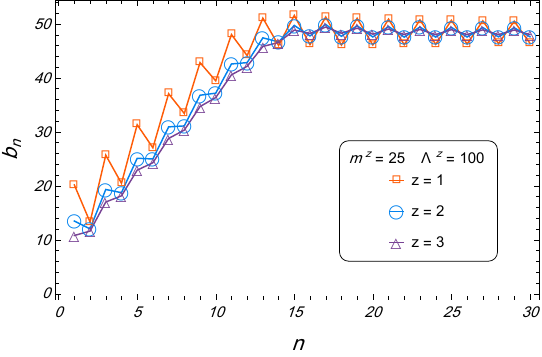}
	\caption{Lanczos coefficients for different values of $z$ in $d=7$ with a hard UV cutoff.}
	\label{fig:UVLC2}
\end{figure}

\begin{figure}[H]
	\centering
	\includegraphics[width=0.49\linewidth]{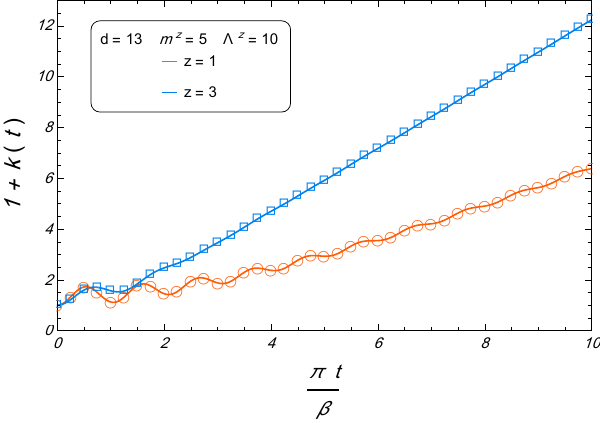}
	\hspace*{.01cm}
	\includegraphics[width=0.49\linewidth]{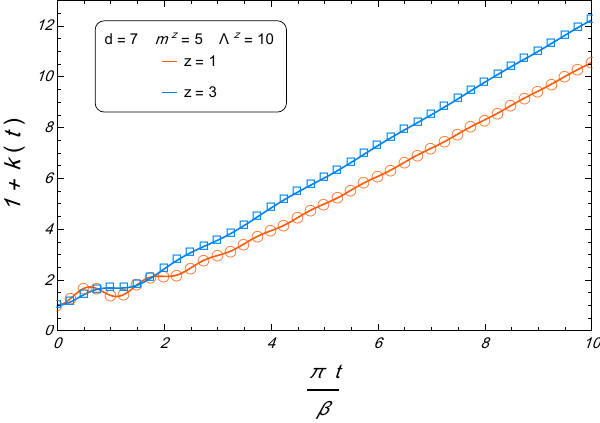}
	\caption{K-complexity in the presence of a hard UV cutoff for various $z$ values.}
	\label{fig:KCm50m100}
\end{figure}

\begin{figure}[H]
	\centering
	\includegraphics[width=0.49\linewidth]{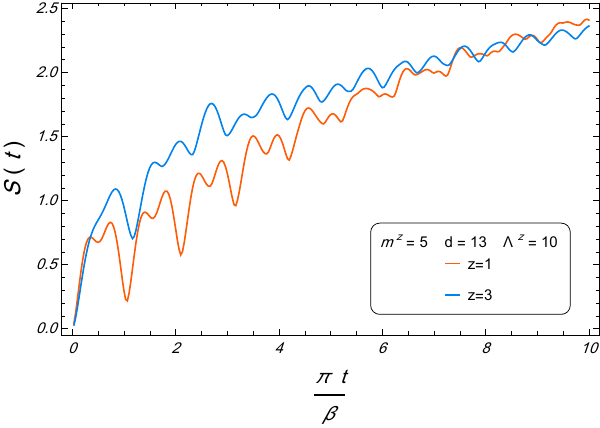}
	\hspace*{.01cm}
	\includegraphics[width=0.49\linewidth]{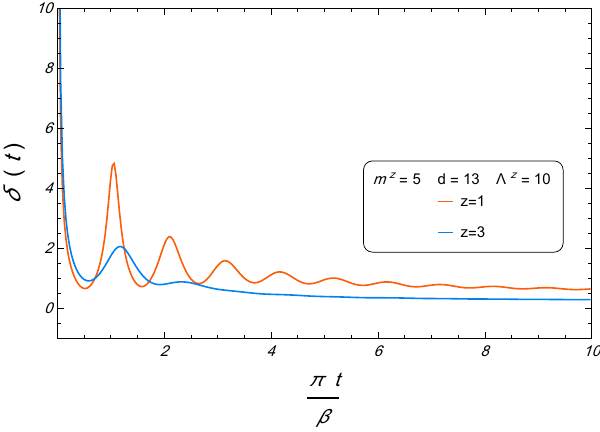}
	\caption{K-entropy and K-variance in the presence of a hard UV cutoff for different $z$ with $\beta=1$.}
	\label{fig:kekvcut}
\end{figure}

The numerical results for different values of $z$ at finite $\Lambda$ are displayed in Fig.~\ref{fig:UVLC2}.This figure shows that the spacing between the Lanczos coefficients depends on $z$, and one can show that the saturation value increases with the UV cutoff.
 The saturation value can be estimated by
\begin{equation}
	b_{\rm s} \approx \frac{\Lambda^z \pm m^z \mp |a_n|}{2} = \frac{\Lambda^z}{2},
\end{equation}
where the signs are taken as $+(-)$ for odd (even) $n$. Thus, the saturation value $b_s$ depends entirely on the UV cutoff $\Lambda$. Moreover, the transition occurs at a saturation point $n=n_s$, estimated by~\cite{Vasli:2023}
\begin{equation}
	n_s \approx \frac{\Lambda^z}{2\alpha},
\end{equation}
which matches well with numerical results. For larger $z$, both $b_s$ and $n_s$ increase.
In this scenario, K-complexity initially exhibits exponential growth, eventually transitioning into a linear regime at late times. As shown in Fig.~\ref{fig:KCm50m100}, for $z>1$, oscillations damp out faster, leading to a cleaner linear behavior compared to the $z=1$ case.
It is also insightful to compute the K-entropy and K-variance when a hard UV cutoff is introduced. Figure~\ref{fig:kekvcut} illustrates these quantities for different $z$. Both exhibit oscillatory behavior, originating from the nonzero mass in the Dirac field.

\subsection{K-complexity for the Lifshitz harmonic model}\label{LCinLHM}
In this section, we study the discretized version of the fermionic model with a finite lattice spacing $a$, leading to a UV cutoff $\Lambda \sim 1/a$. Although Eqs.~\eqref{dispersion_relation_bistar} are not identical, their lattice implementations become equivalent. Consequently, the relations required for even and odd $z$ values are the same, and one can use a unified treatment for both cases.

\begin{figure}[H]
	\centering
	\includegraphics[width=0.49\linewidth]{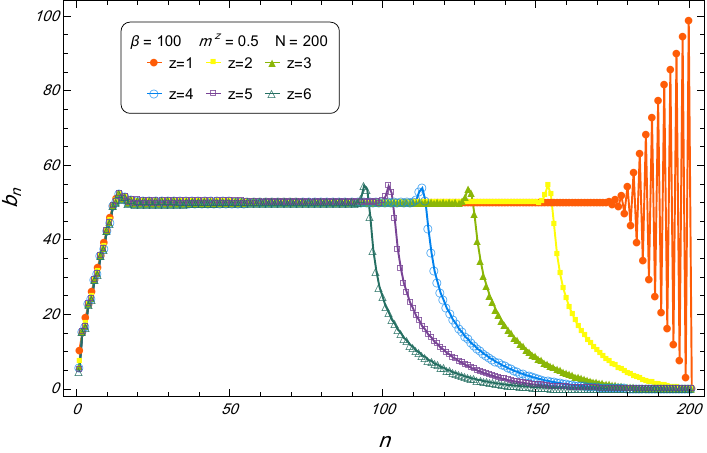}
	\hspace*{.1cm}
	\includegraphics[width=0.49\linewidth]{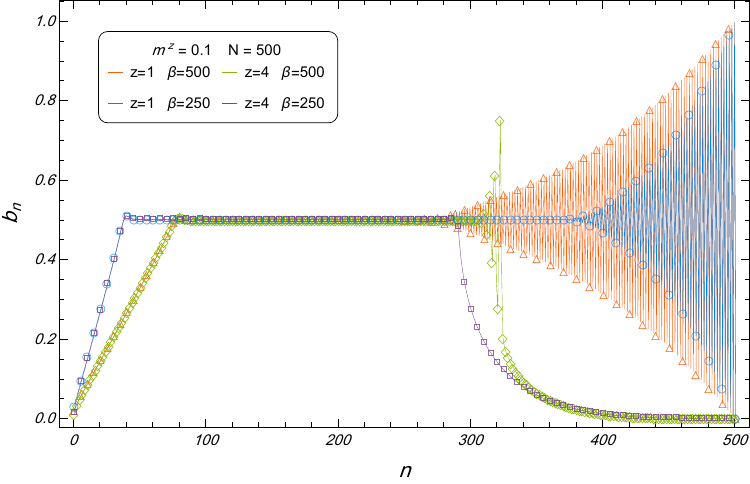}
	\caption{Lanczos coefficients in the Lifshitz harmonic model for different model parameters.}
	\label{fig:LClattice1}
\end{figure}

\begin{figure}[H]
	\centering
	\includegraphics[width=0.45\linewidth]{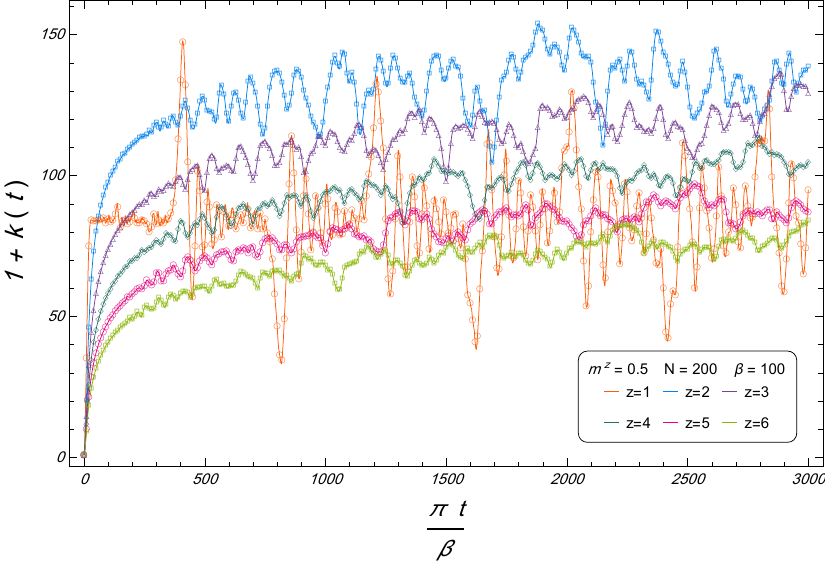}
	\hspace*{.1cm}
	\includegraphics[width=0.45\linewidth]{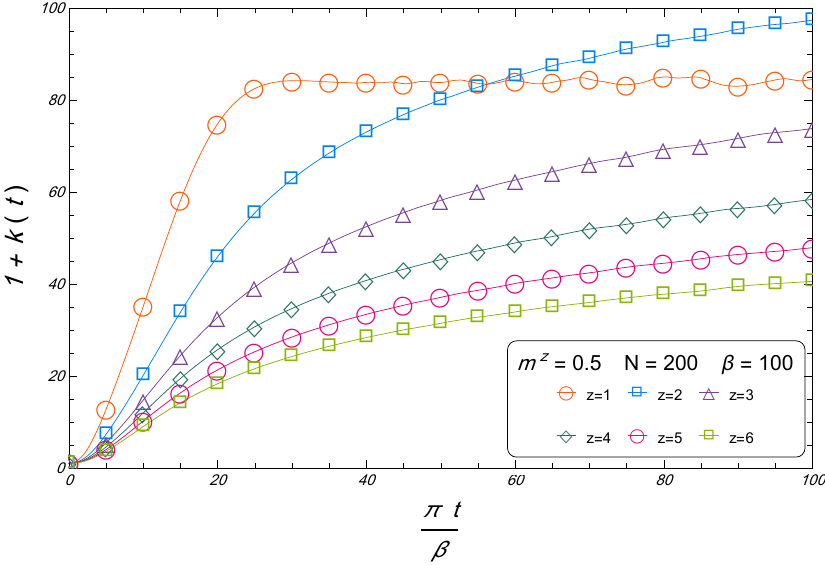}
	\caption{K-complexity in the $f_{\mathrm{I}}(k)$ and $f_{\mathrm{II}}(k)$ models. The right panel zooms on early-time behavior.}
	\label{fig:lattice measures even and odd mark 2}
\end{figure}

\begin{figure}[H]
	\centering
	\includegraphics[width=0.8\linewidth]{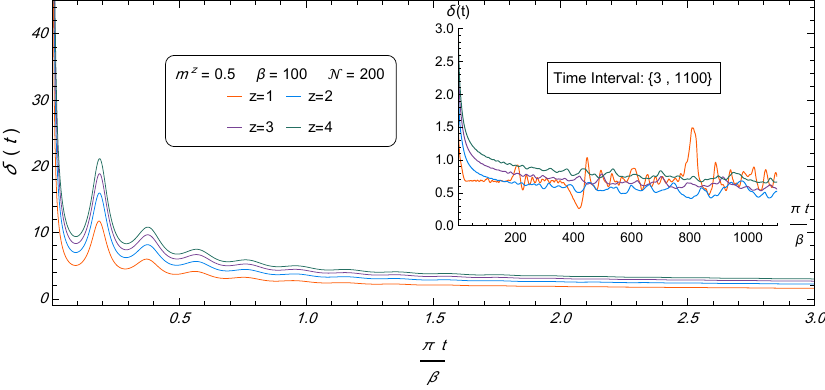}
	\caption{Evolution of K-variance for various $z$ in the $f_{\mathrm{II}}(k)$ model.}
	\label{fig:Kvariance}
\end{figure}

\begin{figure}[H]
	\centering
	\includegraphics[width=0.45\linewidth]{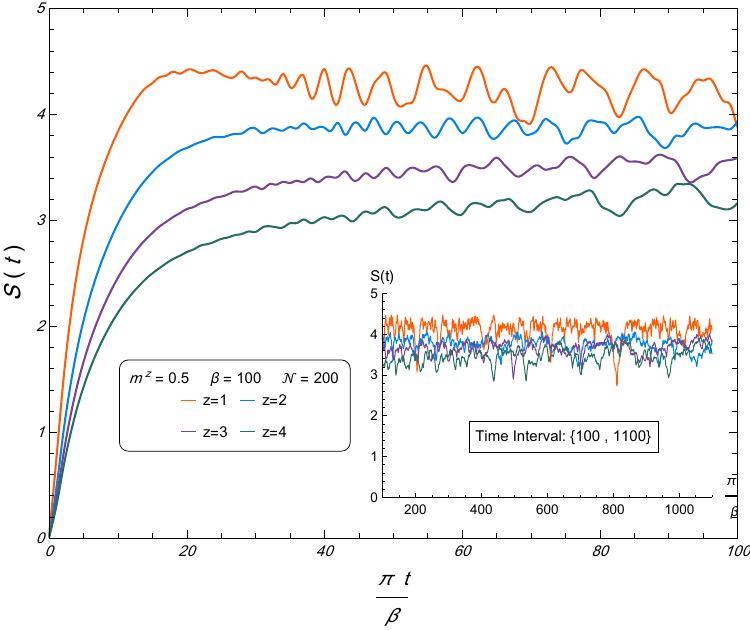}
	\caption{Evolution of K-entropy for various $z$ in the $f_{\mathrm{II}}(k)$ model.}
	\label{fig:Kentropy}
\end{figure}

A modified approach introduced in~\cite{Vasli:2023} simplifies and accelerates the numerical computations required for studying operator growth in lattice models. This method improves the efficiency of extracting K-complexity from operator evolution in discrete systems. Starting from Eq.~\eqref{WPS} and inserting it into Eq.~\eqref{norm}, reordering the integration gives the normalization factor:
\begin{equation}\label{norm1}
	\mathcal{N}^{-1} = \int_0^{\infty} dk \frac{k^{d-2}}{\cosh \left( \frac{\beta \epsilon_k}{2} \right)}.
\end{equation}
Next, combining Eqs.~\eqref{mu2n} and~\eqref{WPS} yields the moments:
\begin{equation}\label{mu2n1}
	\mu_n = \mathcal{N} \int_0^{\infty} dk\, k^{d-2}\, \frac{-m^z\,\epsilon_k^{n-1}}{\cosh\left(\frac{\beta \epsilon_k}{2}\right)}.
\end{equation}
The autocorrelation function $\phi_0(t)$ is obtained via the inverse Fourier transform:
\begin{equation}\label{phi0}
	\phi_0(t) = \mathcal{N} \int_0^{\infty} dk \, k^{d-2} \frac{\cos (\epsilon_k t) + i m^z \sin (\epsilon_k t)}{\cosh \left( \frac{\beta \epsilon_k}{2} \right)}.
\end{equation}
To compute K-complexity numerically, the $n$-th derivative of $\phi_0(t)$ is required, given by
\begin{equation}\label{phi0nth}
	\frac{\partial^n \phi_0(t)}{\partial t^n} = \mathcal{N} \int_{0}^{\infty} dk \; k^{d-2} \; \frac{\epsilon_k^n \cos\left( \frac{\pi n}{2} + t \epsilon_k \right) + i m^z \epsilon_k^n \sin\left( \frac{\pi n}{2} + t \epsilon_k \right)}{\cosh\left( \frac{\beta \epsilon_k}{2} \right)}.
\end{equation}
The lattice dispersion relation is
\begin{equation}\label{dispersionrelationIandII}
	\epsilon_k = \sqrt{\sin^{2z}\left( \frac{\pi k}{N} \right) + m^{2z}},
\end{equation}
where $k$ labels the discrete momenta on a lattice of $N$ sites. Due to the finite lattice size, the Krylov basis becomes finite-dimensional. According to Favard's theorem, if the spectral measure has finite support, the Krylov recursion terminates after finitely many steps~\cite{MuckYang2022}. The lattice version of the Wightman power spectrum becomes
\begin{equation}\label{fwlattice}
	f^W(\omega) = \mathcal{N} \sum_{k=1}^N \frac{\omega - m^z}{\cosh\left( \frac{\beta \omega}{2} \right)} \frac{1}{\epsilon_k} \left[ \delta(\omega - \epsilon_k) - \delta(\omega + \epsilon_k) \right].
\end{equation}
Now the moments on the lattice are computed as
\begin{equation}\label{mioI}
	\mu_n =
	\begin{cases}
		\displaystyle \frac{\mathcal{N}}{\pi} \sum\limits_{k=1}^N (\epsilon_k)^n \, \text{sech}\left( \frac{\beta \epsilon_k}{2} \right), & \text{even } n, \\[2mm]
		\displaystyle \frac{\mathcal{N}}{\pi} \sum\limits_{k=1}^N (\epsilon_k)^{n-1} m^z \, \text{sech}\left( \frac{\beta \epsilon_k}{2} \right), & \text{odd } n.
	\end{cases}
\end{equation}
This discretized structure alters the long-time behavior of complexity compared to the continuum case, with the finite lattice introducing a natural saturation. Substituting Eq.~\eqref{mioI} into Eq.~\eqref{recursion} enables the numerical computation of the Lanczos coefficients. The results for different $\beta$, $m$, $N$, and $z$ values are shown in Fig.~\ref{fig:LClattice1}.
Figure~\ref{fig:LClattice1} shows that in the lattice model, the Lanczos coefficients initially resemble continuum behavior but eventually saturate and then vanish. The finiteness of $N$ restricts Krylov basis growth, in contrast to the continuum where it can grow indefinitely. In lattice models, we focus on the evolution of a local operator, not the full Hilbert space. The Krylov space has dimension $\sim N$, much smaller than the full Hilbert space dimension $2^N$~\cite{Camargo:2022}. For large $N$, the early Lanczos coefficients mimic continuum behavior before saturating.
Interestingly, the structure of the Lanczos coefficients (initial rise, saturation, and vanishing) resembles the findings of~\cite{Barbon:2019BeyondScrambling}, even though the underlying frameworks differ. While no analytical connection is yet established, this similarity hints at possible universality in late-time operator dynamics across finite-dimensional systems. To compute K-complexity in the lattice setup, we use
\begin{equation}\label{phimark1}
	\phi_0(t) = \frac{\mathcal{N}}{\pi} \sum_{k=1}^{N} \sech\left( \frac{1}{2} \beta \epsilon_k \right) 
	\left( \cos (t \epsilon_k) + i \frac{m^z}{\epsilon_k} \sin (t \epsilon_k) \right),
\end{equation}
and insert into Eq.~\eqref{KC}. The numerical results for K-complexity are shown in Fig.~\ref{fig:lattice measures even and odd mark 2}.\footnote{In the lattice model, the total probability sums exactly to one due to the finite-dimensional Krylov space. In continuum cases, only approximate normalization is achieved numerically.}
Figure~\ref{fig:lattice measures even and odd mark 2} shows that K-complexity saturates due to the finite lattice size. Increasing $z$ suppresses K-complexity, as stronger anisotropy restricts operator spreading. K-variance and K-entropy evolution are shown in Figs.~\ref{fig:Kvariance} and~\ref{fig:Kentropy}.
Figure~\ref{fig:Kvariance} shows that lower $z$ leads to smaller K-variance, implying more uniform operator spreading, while higher $z$ leads to broader variance and localization.
Similarly, Fig.~\ref{fig:Kentropy} shows that K-entropy initially grows exponentially before saturating due to the finite Krylov dimension. Again, higher $z$ suppresses entropy growth, consistent with reduced complexity and operator delocalization. Overall, these results highlight the distinctive features of K-complexity in discrete lattice systems compared to continuum field theories.

\section{Conclusions}\label{Conclusions}

In this work, we investigated the K-complexity of Dirac field theories with Lifshitz scaling. By extending the relativistic Dirac model to include a dynamical critical exponent $z$, we explored how anisotropic scaling affects operator growth and information scrambling in both continuum and lattice settings. Below is a concise summary of our main findings:

\begin{itemize}
	\item \textbf{Massless Dirac Field:}
	
	For the massless theory, we observed that the growth of Lanczos coefficients $b_n$ slows down with increasing $z$, reflecting a diminished rate of operator spreading in Krylov space. The scaling parameter $\xi = (d-1)/z$ emerged as a key quantity, unifying the behavior of K-complexity across different spacetime dimensions $d$ and Lifshitz exponents $z$. K-entropy and K-variance analyses confirmed that operator delocalization increases over time, although higher $z$ suppresses this effect.
	
	\item \textbf{Massive Dirac Field:}
	
	Introducing a finite mass $m$ opens a spectral gap and modifies both even and odd moments of the power spectrum. While early-time exponential growth in K-complexity persists, larger masses suppress operator spreading. The alternating structure of the Krylov recursion coefficients $(a_n, b_n)$ arises from intrinsic fermionic features. Larger $z$ further reduces complexity growth, emphasizing the role of anisotropic scaling in limiting operator spreading.
	
	\item \textbf{Effects of a UV Cutoff:}
	
	Introducing a hard UV cutoff $\Lambda$ modifies the growth of K-complexity in two stages. Initially, K-complexity exhibits rapid (often exponential-like) growth. Beyond a characteristic timescale tied to $\Lambda$, the Lanczos coefficients saturate at $b_s \sim \Lambda^z/2$, causing K-complexity to transition from exponential to linear growth at late times. Importantly, in the continuum (with only a momentum cutoff but infinite degrees of freedom), K-complexity grows linearly without saturating. Thus, the UV cutoff constrains the rate but not the indefinite spread of complexity.
	
	\item \textbf{Lattice Realization:}
	
	In the discretized lattice model of the Dirac field, the finite number of momentum modes imposes a fundamental bound on Krylov space, leading to saturation of K-complexity at late times. This behavior differs markedly from the continuum case. For $z > 1$, increasing the inverse temperature $\beta$ delays saturation onset, while larger $z$ accelerates it. These effects highlight distinct operator dynamics in lattice quantum field theories compared to their continuum counterparts.
\end{itemize}

In the lattice formulation, the Krylov basis is finite due to discretization. As evident from Eq.~\eqref{phimark1}, in the zero-temperature limit ($\beta \to \infty$), the autocorrelation function vanishes, $\phi_0(t) = 0$, implying that K-complexity remains zero at all times. This reflects the suppression of operator spreading in the absence of thermal excitations.

\begin{figure}[H]
	\centering
	\includegraphics[height=4.9cm]{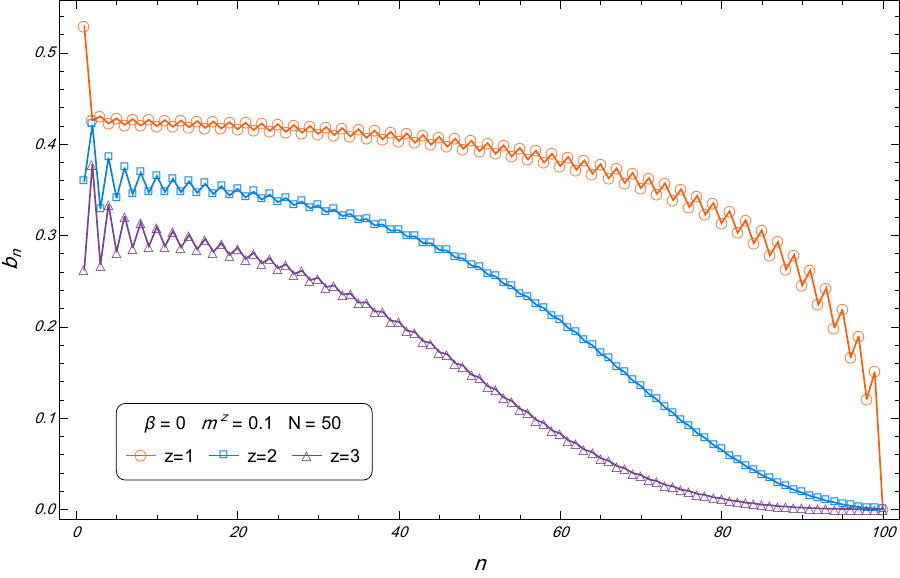}
	\hspace*{.1cm}
	\includegraphics[height=5cm]{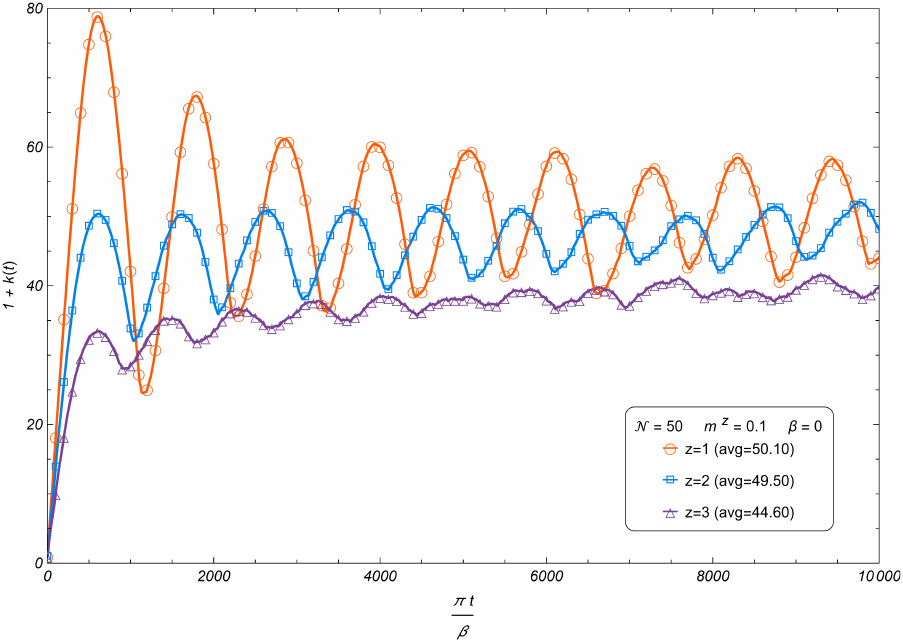}
	\caption{K-complexity in the lattice Dirac model at $\beta = 0$ for various $z$. Strong early-time operator spreading is observed at $z=1$, while increasing $z$ leads to gradual suppression of complexity growth. (Plots cover $t = 0$ to $t = 200{,}000$.)}
	\label{fig:betazeromassivemasslessdiraclattice}
\end{figure}

In contrast, at infinite temperature ($\beta = 0$), numerical computations reveal strong operator spreading, particularly at $z=1$. In this regime, K-complexity exhibits significant early-time growth\footnote{It is worth mentioning that although early-time peaks in K-complexity have been discussed as potential signatures of chaotic dynamics (see, e.g., Refs.~\cite{Baggioli:2024KCP,Alishahiha:2024ChaosProbe,Camargo:2024SpinChains,Huh:2024MixedPhase,Erdmenger:2023Universal,Balasubramanian:2022Spread}), our results do not exhibit such peaks.}, accompanied by strong fluctuations around the mean value. The average K-complexity scales proportionally to the number of lattice sites $N$, reflecting extensive operator exploration across the Krylov basis. As illustrated in Fig.~\ref{fig:betazeromassivemasslessdiraclattice}, increasing the dynamical exponent $z$ leads to a gradual suppression of operator spreading. However, as shown in Fig.~\ref{fig:lattice measures even and odd mark 2}, for finite $\beta$ (e.g., $\beta = 100$), operator spreading is noticeably suppressed, and the operator does not fully explore the Krylov basis. Increasing $z>1$ further enhances this suppression, leading to more constrained operator dynamics.

An interesting distinction emerged regarding the thermal structures of K-complexity in bosonic and fermionic Lifshitz-type theories. In the scalar field case, the Wightman function involves a thermal factor $1/\sinh(\beta\omega/2)$~\cite{Vasli:2023}, whereas for the Dirac field, it involves $1/\cosh(\beta\omega/2)$. This distinction stems from quantum statistics: bosonic fields are periodic, and fermionic fields are antiperiodic in Euclidean time, leading to different Matsubara frequencies and thermal occupations.
Physically, this explains why the Dirac field yields finite, well-behaved K-complexity at $\beta \to 0$, owing to Pauli exclusion limiting fermionic occupation numbers. In contrast, bosonic fields allow divergent occupation at high temperatures, resulting in divergent moments and ill-defined Lanczos coefficients. Thus, the finiteness of fermionic K-complexity at high temperatures has a deep statistical origin, sharply distinguishing it from bosonic operator growth, even in free field theories.

Across all cases considered—massless, massive, continuum, and lattice—the temperature plays a central role in the emergence of K-complexity. In the zero-temperature limit, complexity vanishes as the system remains in its ground state. Conversely, in the infinite-temperature limit, complexity is activated and governed by the underlying field statistics and scaling symmetries. We conclude that in free field theories, temperature is essential for activating operator growth, complexity, and the rich dynamical structures studied through Krylov diagnostics.

\section*{Acknowledgments}
We would like to thank Mohammad Javad vasli for fruitful discussions. We are grateful to Mohsen Alishahiha for careful reading of the manuscript and his valuable comments. The work of M. Reza Mohammadi Mozaffar is supported by the Iran National Science Foundation (INSF) under project No. 4036941.

\appendix
\section{Additional Calculations}
\label{appendix:calculations}

We have defined the shorthand functions for Eq.~\ref{miok} as follows

\begingroup
\allowdisplaybreaks
\small
\begin{align*}
	A_1^{(\text{even})}(n,d,z,m,\beta) &= -\frac{2^{\frac{2}{z}+2}\,\csc\!\left(\frac{\pi(d+2nz-1)}{2z}\right)(\beta m^z)^{\frac{d+2nz-1}{z}}\,\mathcal{F}_1^{(\text{even})}(n)}
	{\Gamma(-n)}, \\[2mm]
	A_2^{(\text{even})}(n,d,z,m,\beta) &= \frac{(e^{2\pi i n}-1)\,4^{1/z}\,\csc\!\left(\frac{\pi(d+2nz-1)}{2z}\right)(\beta m^z)^{\frac{d+2nz+z-1}{z}}\,\mathcal{F}_2^{(\text{even})}(n)}
	{\Gamma(-n)}, \\[2mm]
	A_3^{(\text{even})}(n,d,z,m,\beta) &= \frac{e^{i\pi n}\,4^{\frac{d}{z}+2n+1}\,\csc\!\left(\frac{\pi(d+2nz-1)}{z}\right)}
	{\Gamma\left(\frac{d-1}{2z}\right)}
	\left[ 4\cos(\pi n)\,\mathcal{F}_3^{(\text{even})}(n) - i\beta m^z\sin(\pi n)\,\mathcal{F}_4^{(\text{even})}(n) \right], \\[2mm]
	A_4^{(\text{even})}(n,d,z,m,\beta) &= -\frac{4^{1/z}\,\sec\!\left(\frac{\pi(d+2nz-1)}{2z}\right)(\beta m^z)^{\frac{d+2nz-1}{z}}}
	{\Gamma\left(\frac{1}{2}-n\right)}
	\left[ \beta(2n+1)m^z\cos(\pi n)\,\mathcal{F}_5^{(\text{even})}(n) - 8i\sin(\pi n)\,\mathcal{F}_6^{(\text{even})}(n) \right], \\[2mm]
	\mathcal{F}_1^{(\text{even})}(n) &= {}_1\tilde{F}_2\left(n+1;\, \tfrac{1}{2},\, \tfrac{d-1}{2z}+n+1;\, \tfrac{m^{2z}\beta^2}{16}\right), \\[1mm]
	\mathcal{F}_2^{(\text{even})}(n) &= {}_1\tilde{F}_2\left(n+1;\, \tfrac{3}{2},\, \tfrac{d-1}{2z}+n+1;\, \tfrac{m^{2z}\beta^2}{16}\right), \\[1mm]
	\mathcal{F}_3^{(\text{even})}(n) &= {}_1\tilde{F}_2\left(1-\tfrac{d-1}{2z};\, \tfrac{-d-2nz+z+1}{2z},\, -\tfrac{d-1}{2z}-n+1;\, \tfrac{m^{2z}\beta^2}{16}\right), \\[1mm]
	\mathcal{F}_4^{(\text{even})}(n) &= {}_1\tilde{F}_2\left(1-\tfrac{d-1}{2z};\, -\tfrac{d-1}{2z}-n+1,\,-\tfrac{d+2nz-3z-1}{2z};\, \tfrac{m^{2z}\beta^2}{16}\right), \\[1mm]
	\mathcal{F}_5^{(\text{even})}(n) &= {}_1\tilde{F}_2\left(n+\tfrac{3}{2};\, \tfrac{3}{2},\, \tfrac{d+2nz+3z-1}{2z};\, \tfrac{m^{2z}\beta^2}{16}\right), \\[1mm]
	\mathcal{F}_6^{(\text{even})}(n) &= {}_1\tilde{F}_2\left(n+\tfrac{1}{2};\, \tfrac{1}{2},\, \tfrac{d+2nz+z-1}{2z};\, \tfrac{m^{2z}\beta^2}{16}\right).
\end{align*}
\endgroup
We rewrite Eq.~\ref{miok} in terms of the scaling parameter 
\[
\xi=\frac{d-1}{z}\quad\Longrightarrow\quad d=\xi z+1,
\]
so that all dependence on $d$ and $z$ is encoded in $\xi$. Consequently, any pair $(d,z)$ with the same \(\xi\) (for example, $(6,1)$, $(11,2)$, $(16,3)$, etc. for $\xi=5$) will yield identical moments and Lanczos coefficients.
\begingroup
\allowdisplaybreaks
\small
\begin{align*}
	\mu_{\text{even}} &= \frac{\pi^2\,2^{-\xi-3} e^{i\pi n}\, \tilde{m}^{-\frac{1}{2}(\xi+1)} \, \beta^{\frac{1}{2}(-\xi-4n-1)}}
	{K_{\frac{\xi+1}{2}}\left( \frac{\tilde{m}\beta}{2} \right)} \\[1mm]
	&\quad \times \Biggl\{ 
	\frac{(-1)^\xi\,2^{2\xi+4n+1}}{\Gamma\left( \frac{\xi}{2} \right)}
	\Biggl[4\,\csc(\pi n)\, {}_1\tilde{F}_2\left(1-\frac{\xi}{2};\,-\frac{\xi}{2}-n+\tfrac{1}{2},\,-\frac{\xi}{2}-n+1;\,\frac{(\tilde{m})^2\beta^2}{16}\right) \\[1mm]
	&\qquad\quad - i\,\beta\,\tilde{m}\,\sec(\pi n)\, {}_1\tilde{F}_2\left(1-\frac{\xi}{2};\,-\frac{\xi}{2}-n+1,\,-\frac{\xi}{2}-n+\tfrac{3}{2};\,\frac{(\tilde{m})^2\beta^2}{16}\right)
	\Biggr] \\[2mm]
	&\quad - \frac{2\,\csc\left(\frac{\pi \xi}{2}+\pi n\right)(\beta\,\tilde{m})^{\xi+2n}}{\Gamma(-n)}
	\Biggl[4\,\cos(\pi n)\, {}_1\tilde{F}_2\left(n+1;\,\tfrac{1}{2},\,\frac{\xi}{2}+n+1;\,\frac{(\tilde{m})^2\beta^2}{16}\right) \\[1mm]
	&\qquad\quad - i\,\beta\,\tilde{m}\,\sin(\pi n)\, {}_1\tilde{F}_2\left(n+1;\,\tfrac{3}{2},\,\frac{\xi}{2}+n+1;\,\frac{(\tilde{m})^2\beta^2}{16}\right)
	\Biggr] \\[2mm]
	&\quad - \frac{(\beta\,\tilde{m})^{\xi+2n}\,\sec\left(\frac{\pi \xi}{2}+\pi n\right)}{\Gamma\left( \frac{1}{2} - n \right)}
	\Biggl[\beta(2n+1)\,\tilde{m}\,\cos(\pi n)\, {}_1\tilde{F}_2\left(n+\tfrac{3}{2};\,\tfrac{3}{2},\,\frac{\xi+3}{2}+n;\,\frac{(\tilde{m})^2\beta^2}{16}\right) \\[1mm]
	&\qquad\quad - 8i\,\sin(\pi n)\, {}_1\tilde{F}_2\left(n+\tfrac{1}{2};\,\tfrac{1}{2},\,\frac{\xi+1}{2}+n;\,\frac{(\tilde{m})^2\beta^2}{16}\right)
	\Biggr]
	\Biggr\}.
\end{align*}
\endgroup
We have also defined the shorthand functions for Eq.~\ref{mioodd} as follows
\begingroup
\allowdisplaybreaks
\small
\begin{align*}
	A_1^{(\text{odd})}(n,d,z,m,\beta) &= \frac{4^{1/z}\,\csc\!\left(\frac{\pi(d+2nz-1)}{2z}\right)(\beta m^z)^{\frac{d+2nz+z-1}{z}}\,\mathcal{F}_1^{(\text{odd})}(n)}
	{\Gamma(-n-1)}, \\[2mm]
	A_2^{(\text{odd})}(n,d,z,m,\beta) &= \frac{e^{i\pi n}\,4^{\frac{d}{z}+2n+2}\,\csc\!\left(\frac{\pi(d+2nz-1)}{z}\right)}
	{\Gamma\left(\frac{d-1}{2z}\right)}
	\left[\beta m^z\cos(\pi n)\,\mathcal{F}_2^{(\text{odd})}(n)-4i\sin(\pi n)\,\mathcal{F}_3^{(\text{odd})}(n)\right], \\[2mm]
	A_3^{(\text{odd})}(n,d,z,m,\beta) &= \frac{e^{i\pi n}\,2^{\frac{z+2}{z}}\,\sec\!\left(\frac{\pi(d+2nz-1)}{2z}\right)(\beta m^z)^{\frac{d+2nz+z-1}{z}}}
	{\Gamma\left(-n-\frac{1}{2}\right)}
	\left[\beta m^z\cos(\pi n)\,\mathcal{F}_4^{(\text{odd})}(n)-4i\sin(\pi n)\,\mathcal{F}_5^{(\text{odd})}(n)\right], \\[4mm]
	\mathcal{F}_1^{(\text{odd})}(n) &= -\frac{4\left((-1)^{2n}+1\right)}{n+1}\,{}_{1}\tilde{F}_2\left(n+1;\,\frac{1}{2},\,\frac{d-1}{2z}+n+1;\,\frac{m^{2z}\beta^2}{16}\right) \\[1mm]
	&\quad +\beta\left(-1+e^{2i\pi n}\right)m^z\,{}_{1}\tilde{F}_2\left(n+2;\,\frac{3}{2},\,\frac{d-1}{2z}+n+2;\,\frac{m^{2z}\beta^2}{16}\right), \\[2mm]
	\mathcal{F}_2^{(\text{odd})}(n) &= {}_{1}\tilde{F}_2\left(1-\frac{d-1}{2z};\,\frac{-d-2nz+z+1}{2z},\,-\frac{d-1}{2z}-n+1;\,\frac{m^{2z}\beta^2}{16}\right), \\[2mm]
	\mathcal{F}_3^{(\text{odd})}(n) &= {}_{1}\tilde{F}_2\left(1-\frac{d-1}{2z};\,-\frac{d+2nz-1}{2z},\,\frac{-d-2nz+z+1}{2z};\,\frac{m^{2z}\beta^2}{16}\right), \\[2mm]
	\mathcal{F}_4^{(\text{odd})}(n) &= {}_{1}\tilde{F}_2\left(n+\frac{3}{2};\,\frac{3}{2},\,\frac{d+2nz+3z-1}{2z};\,\frac{m^{2z}\beta^2}{16}\right), \\[2mm]
	\mathcal{F}_5^{(\text{odd})}(n) &= {}_{1}\tilde{F}_2\left(n+\frac{3}{2};\,\frac{1}{2},\,\frac{d+2nz+3z-1}{2z};\,\frac{m^{2z}\beta^2}{16}\right).
\end{align*}
\endgroup
We now rewrite Eq.\refeq{mioodd} in terms of the scaling parameter, as done before, so that all dependence on the parameters is absorbed into a single parameter $\xi = \frac{d-1}{z}$.
\begingroup
\allowdisplaybreaks
\small
\begin{align*}
	\mu_{\text{odd}} &= \frac{\pi^2\,2^{-\xi-3}\,\tilde{m}^{-\frac{1}{2}(\xi+1)}\,\beta^{\frac{1}{2}(-\xi-4n-3)}}
	{K_{\frac{\xi+1}{2}}\left(\frac{\tilde{m}\beta}{2}\right)} \\[1mm]
	&\quad \times \Biggl\{ 
	\frac{\csc\left(\frac{\pi \xi}{2}+\pi n\right)(\beta\,\tilde{m})^{\xi+2n+1}}{\Gamma(-n-1)}
	\Biggl[-\frac{4\left((-1)^{2n}+1\right)}{n+1}\, {}_1\tilde{F}_2\left(n+1;\,\frac{1}{2},\,\frac{\xi}{2}+n+1;\,\frac{(\tilde{m})^2\beta^2}{16}\right) \\[1mm]
	&\qquad\quad +\; \beta\left(-1+e^{2i\pi n}\right)\tilde{m}\, {}_1\tilde{F}_2\left(n+2;\,\frac{3}{2},\,\frac{\xi}{2}+n+2;\,\frac{(\tilde{m})^2\beta^2}{16}\right)
	\Biggr] \\[2mm]
	&\quad +\; \frac{(-1)^\xi e^{i\pi n}\,2^{2\xi+4n+3}}{\Gamma\left(\frac{\xi}{2}\right)}
	\Biggl[-\beta\,\tilde{m}\,\csc(\pi n)\, {}_1\tilde{F}_2\left(1-\frac{\xi}{2};\,-\frac{\xi}{2}-n+\tfrac{1}{2},\,-\frac{\xi}{2}-n+1;\,\frac{(\tilde{m})^2\beta^2}{16}\right) \\[1mm]
	&\qquad\quad +\; 4i\,\sec(\pi n)\, {}_1\tilde{F}_2\left(1-\frac{\xi}{2};\,-\frac{\xi}{2}-n,\,-\frac{\xi}{2}-n+\tfrac{1}{2};\,\frac{(\tilde{m})^2\beta^2}{16}\right)
	\Biggr] \\[2mm]
	&\quad -\; \frac{2(\beta\,\tilde{m})^{\xi+2n+1}\,\sec\left(\frac{\pi \xi}{2}+\pi n\right)}{\Gamma\left(-n-\frac{1}{2}\right)}
	\Biggl[\beta(2n+1)\,\tilde{m}\,\cos(\pi n)\, {}_1\tilde{F}_2\left(n+\tfrac{3}{2};\,\tfrac{3}{2},\,\frac{\xi+3}{2}+n;\,\frac{(\tilde{m})^2\beta^2}{16}\right) \\[1mm]
	&\qquad\quad -\; 8i\,\sin(\pi n)\, {}_1\tilde{F}_2\left(n+\tfrac{1}{2};\,\tfrac{1}{2},\,\frac{\xi+1}{2}+n;\,\frac{(\tilde{m})^2\beta^2}{16}\right)
	\Biggr]
	\Biggr\}.
\end{align*}
\endgroup

\end{document}